\documentclass[reprint,aps,amsmath,amssymb,notitlepage,tightenlines,pra,unsortedaddress]{revtex4-1}
\usepackage{bm}
\usepackage{amsmath}
\usepackage{amssymb}
\usepackage{amsfonts}
\usepackage{graphicx}
\usepackage{color}
\usepackage[caption=false]{subfig}
\usepackage{cleveref}

\definecolor{violet}{rgb}{1,0,1}
\definecolor{dgreen}{rgb}{0,.5,0}
\definecolor{dred}{rgb}{0.5,0,0}

\newcommand{\revc}[1]{{\color{black} #1}}

\newcommand{\revcc}[1]{{\color{black} #1}}

\begin{document}

\title{Coupled Maxwell and Time-Dependent Orbital Free Density Functional Calculations}

\author{Cody Covington}
\affiliation{Department of Chemistry, Austin Peay State University, Clarksville, Tennessee, 37043, USA}
\affiliation{Department of Physics and Astronomy, Vanderbilt University, Nashville, Tennessee, 37235, USA}

\author{Justin Malave}
\affiliation{Department of Physics and Astronomy, Vanderbilt University, Nashville, Tennessee, 37235, USA}

\author{K\'alm\'an Varga}
\email{kalman.varga@vanderbilt.edu}
\affiliation{Department of Physics and Astronomy, Vanderbilt University, Nashville, Tennessee, 37235, USA}

\begin{abstract}
Coupled Maxwell and time-dependent orbital-free calculations are implemented and tested
to describe the interaction of electromagnetic waves and matter. The currents and induced fields predicted by the  orbital-free calculations are compared to time-dependent density functional calculations and very good agreement is found for various systems including jellium sheets, jellium spheres, atomistic sheets and icosahedron clusters.
\end{abstract}

\maketitle

\section{\label{sec:intro}Introduction}
Ground state density functional theory (DFT) \cite{PhysRev.136.B864} Kohn-Sham  (KS) \cite{PhysRev.140.A1133} calculations are very successful in condensed matter physics, 
because they provide a reasonably accurate description of material properties in a computationally manageable way. \revcc{Mostly these} DFT calculations are based on KS orbitals and albeit some approaches scale linearly with the number of orbitals \cite{RevModPhys.71.1085,Bowler_2012} for certain systems,
\revcc{in general KS orbital based DFT} has cubic scaling with respect to the system size. Orbital-free (OF) DFT   \cite{Wang2002,Wesolowski2013,Karasiev2014,
witt_delrio_dieterich_carter_2018} became an attractive alternative because its main variable is the electron density and it computationally scales linearly with system size. 
OF-DFT has shown considerable success in million-atom material simulations of metals, 
\cite{witt_delrio_dieterich_carter_2018,SHAO201878,HUNG2009163}, warm dense matter \cite{PhysRevLett.111.175002,PhysRevLett.121.145001}, 
quantum dots \cite{PhysRevB.100.041105,shao2020newgeneration,HO200757,PhysRevB.63.115404,doi:10.1021/jp0018504}. Computer codes for efficient implementation of OF-DFT have also been developed \cite{CHEN2015228,doi:10.1002/wcms.1482,SHAO201878,GOLUB2020107365}.

OF-DFT directly calculates the ground-state electron density
minimizing the energy functional without using KS orbitals. The minimization leads to a single Schr\"odinger-like equation. 
This equation contains a potential, that is derived from the kinetic energy (KE) functional \cite{PhysRevA.44.5521,PhysRevA.90.062515} to enforce the antisymmetry requirement of the many-electron wave function. 

Time-dependent problems, e.g. calculation of electronic stopping power \cite{PhysRevB.98.144302,PhysRevLett.121.145001}, spectra of clusters \cite{doi:10.1002/wcms.1482}, plasmonics \cite{doi:10.1021/acs.jpcc.6b05841,
doi:10.1021/acs.jpcc.9b10510,doi:10.1021/jz500216t} have been also studied with the "time-dependent" OF (TD-OF) approach. In this case a single time-dependent Schr\"odinger equation has to be solved with the additional difficulty that the KE potential is time-dependent \cite{doi:10.1063/1.3574347}.

The main difficulty of the OF calculations is \revcc{finding} the appropriate approximations for the kinetic energy functional.
The simplest approximations use local or semilocal expressions, where the energy density depends only on the density and its derivatives at one spatial point \cite{PhysRevA.38.625,PhysRevB.86.115101,PhysRevB.98.041111,doi:10.1021/acs.jctc.9b00183,doi:10.1021/acs.jpclett.8b01926,PhysRevB.100.165111}.
\revc{However,} the local and semilocal KE functionals do not have the correct linear response properties. To correct this, 
nonlocal (or two-point) KE functionals were developed invoking the linear response of the noninteracting homogeneous electron gas
\cite{PhysRevB.49.5220,PhysRevB.45.13196,Perrot_1994}. These nonlocal KE functionals 
work well for \revc{a} nearly free electron gas and \revc{were} developed further for more complicated cases \cite{PhysRevB.32.7868,PhysRevB.58.13465,PhysRevB.60.16350,PhysRevB.86.235109}. These functionals are computationally expensive because they depend explicitly on density at two spatially separated points. More recently, efficient alternatives based on a line integral representation of the KE functional have been proposed and tested \cite{doi:10.1063/1.5023926,PhysRevB.100.205132,PhysRevB.100.041105}. 
\revc{While} nonlocal functionals are considered to be superior, semilocal functionals are faster and new semilocal functionals have been developed that show very good accuracy \cite{doi:10.1021/acs.jpclett.8b01926,PhysRevB.100.165111}. Recent advances in machine learning techniques were also put to work in quest of better KE functionals
\cite{GOLUB2020107365,SEINO2019136732,doi:10.1021/acs.jctc.0c00580,FUJINAMI2020137358,doi:10.1063/5.0015042}.

 The development of dynamic KE functionals for time-dependent problems
 is less advanced than the static ones. In Ref. \cite{doi:10.1063/1.3574347}, a dynamic KE potential is proposed that is time propagated and forces the system to closely fit the susceptibility of the noninteracting homegeneous electron gas (Lindhard function) \cite{osti_4405425}. In Ref. 
\cite{PhysRevB.98.144302} a local current density dynamic KE potential is derived by perturbation theory.  This approach is computationally more efficient because
it does not require the time propagation of the KE potential. Both approaches seem to reproduce the frequency dependent Lindhard function nicely.

In this work we use the TD-OF approach coupled with the Maxwell equations to describe the interaction of electromagnetic waves and matter. In a previous paper \cite{PhysRevE.100.053301} we developed a method to solve the coupled Maxwell TD-KS equations. 
In this coupled frame, the densities and current are calculated in a quantum mechanical framework in the presence of a time-dependent vector
potential and then the Maxwell equations are solved with the calculated time-dependent microscopic currents and densities to obtain the new vector potential.
\revc{In order for the coupling of the quantum and Maxwell dynamics to be significant} one needs large systems with many electrons. One can use  Time Dependent Density Functional Theory (TDDFT) \cite{DFT_PhysRevLett.52.997} (as we did in Ref. \cite{PhysRevE.100.053301}), but that limits the application to relatively small systems. The OF approach allows the simulation of much larger systems (on the scale of millions of atoms 
\cite{witt_delrio_dieterich_carter_2018,SHAO201878,HUNG2009163,PhysRevLett.111.175002,PhysRevLett.121.145001,PhysRevB.100.041105,shao2020newgeneration}).
\revcc{The OF approach may allow for electrodynamic simulations to be performed for systems of relevant size with a quantum description of the electrons.}

The goal of this paper is to implement coupled Maxwell TD-OF calculations and test them against TDDFT results. We will solve the TD-OF and TDDFT equations using a \revc{real--space grid and real--time propagation} \cite{varga_driscoll_2011}. The Maxwell-equations are solved using the Riemann-Silberstein formalism \cite{doi:10.1080/00018732.2019.1695875,PhysRevE.100.053301}. To make the TD-OF 
and TDDFT calculations comparable we add a constraining potential to the TD-OF Hamiltonian.
This potential ensures that the initial ground state densities are the same in the OF and DFT
calculations.

\section{Formalism}
\subsection{Time-dependent Kohn-Sham equations}
 The time-dependent Kohn-Sham (TD-KS) equation for the $i$ the electron orbital is
\begin{equation} 
\left(i \hbar \frac{\partial}{\partial t}-H_{\mathrm{KS}}\right) \psi_{i}(\mathbf{r}, t)=0,
\label{tdks}
\end{equation}
where 
\begin{equation}
H_{\mathrm{KS}}(\mathbf{r}, t)=-\frac{1}{2 m}\left[-i \hbar \nabla_{\mathbf{r}}+\mathbf{A}(\mathbf{r}, t)\right]^{2} 
+V_{\mathrm{KS}}(\mathbf{r}, t).
\end{equation}
The Kohn-Sham potential,
\begin{equation}
V_{\mathrm{KS}}(\mathbf{r}, t)=
V_{\mathrm{ext}}(\mathbf{r})+V_{\mathrm{H}}[\rho](\mathbf{r})+V_{\mathrm{XC}}[\rho](\mathbf{r}) 
\end{equation}
is a sum of the external potential, the Hartree and the exchange correlation terms. The vector potential, 
$\mathbf{A}(\mathbf{r}, t)$ describes the electromagnetic fields.
\revcc{The TD-KS equations can be solved by time propagation} and one can calculate the electron density and current
at any time $t$ as
\begin{equation}
\rho(\mathbf{r}, t)=2 \sum_{i=1}^{N_{e} / 2}\left|\psi_{i}(\mathbf{r}, t)\right|^{2} \\
\end{equation}
\begin{equation}
\mathbf{J}(\mathbf{r}, t)=2 \operatorname{Re} \sum_{i=1}^{N_{e} / 2}\left[\psi_{i}(\mathbf{r}, t)^{*}\left(-i \hbar \nabla_{\mathbf{r}}+\mathbf{A}(\mathbf{r}, t)\right) \psi_{i}(\mathbf{r}, t)\right]
\end{equation}
where $N_{e}$ is the number of electrons in the system and each orbital is occupied by two electrons.
\revc{This current is input into the Maxwell equations, which will then propagate} a vector potential
$\mathbf{A}(\mathbf{r}, t)$.

\subsection{Time-dependent Orbital Free equations}
 The time-dependent orbital free (TD-OF) equation is
\begin{equation} 
\left(i \hbar \frac{\partial}{\partial t}-H_{\mathrm{OF}}\right) \Psi(\mathbf{r}, t)=0,
\label{tdof}
\end{equation}
where 
\begin{equation}
H_{\mathrm{OF}}(\mathbf{r},t)
=-\frac{1}{2 m}\left[-i \hbar \nabla_{\mathbf{r}}+\mathbf{A}(\mathbf{r}, t)\right]^{2} 
+V_{\mathrm{OF}}(\mathbf{r},t).
\label{tdof_h}
\end{equation}
The orbital free potential is defined as
\begin{equation}
V_{\mathrm{OF}}(\mathbf{r}, t)=
V_{\mathrm{KS}}(\mathbf{r})+V_{\mathrm{TF}}[\rho](\mathbf{r})+(a-1)V_{\mathrm{W}}[\rho](\mathbf{r})+V_{\mathrm{c}}(\mathbf{r}),
\end{equation}
where \revcc{$V_{\mathrm{TF}}$ is the Thomas-Fermi kinetic energy functional}, $V_{\mathrm{W}}$ is the von Weizs\"acker potential and $V_\mathrm{c}$ is a constraining potential that we will define later. $a$ is a numerical coefficient of the Weizs\"acker term. Normally the  Weizs\"acker term comes with an "$a$" 
multiplier (with values $a$=1 to 1/9). Here $a-1$ appears because the one has to compensate the appropriate part in the kinetic energy in Eq. \eqref{tdof}.  In this case 
the electron density and current at any time $t$ is
\begin{equation}
\rho_{OF}(\mathbf{r}, t)=\left|\Psi(\mathbf{r}, t)\right|^{2} \\
\end{equation}
\begin{equation}
\mathbf{J}_{OF}(\mathbf{r}, t)=2 \operatorname{Re} \left[\Psi(\mathbf{r}, t)^{*}\left(-i \hbar \nabla_{\mathbf{r}}+\mathbf{A}(\mathbf{r}, t)\right) \Psi(\mathbf{r}, t)\right]
\end{equation}

\subsection{Constrained minimization}
To compare TDDFT and OF one ideally would use the same ground state density, but $V_{KS}$ and $V_{OF}$ produce different
densities, $\rho(\mathbf{r})$ and $\rho_{OF}(\mathbf{r})$, respectively. In Ref. \cite{PhysRevB.98.075108} we have presented a constrained density functional approach. This method can be used to generate a constraining potential that forces the charge density to be equal to a prescribed
density. In the present case we will use the constraint
\begin{equation}
\left|\Psi(\mathbf{r})\right|^{2} =\rho(\mathbf{r})
\label{cons}
\end{equation}
where $\rho(\mathbf{r})$ is the density of the ground state DFT calculation. 

The following iterative procedure can be used to calculate the constraining potential:
\begin{equation}
\Psi^{(n+1)}(\mathbf{r})=\Psi^{(n)}(\mathbf{r})-x_{0}\left(\hat{H}_{\mathrm{OF}}+\lambda^{(n)}\hat{Q}
-\epsilon^{(n)}\right) \Psi^{(n)}(\mathbf{r}), 
\label{cdftiter}
\end{equation}
where $x_0$ controls the convergence, 
\begin{equation}
    \epsilon^{(n)}=\left\langle\Psi^{(n)}|H_\mathrm{OF}| \Psi^{(n)}\right\rangle
\end{equation}
is the energy expectation value, and  $\hat{Q}$ is the density operator such that
\begin{equation}
\left\langle\Psi^{(n)}|\hat{Q}| \Psi^{(n)}\right\rangle=\left|\Psi^{(n)}(\mathbf{r})\right|^{2} .
\end{equation}
Given a desired initial density distribution $\rho(\mathbf{r})$ one looks for the potential $\lambda(\mathbf{r}),$
\begin{equation}
\lambda(\mathbf{r}) \psi^{(n)}(\mathbf{r})=\lambda \hat{Q} \psi^{(n)}
\end{equation}
which constrains the density according to Eq. \eqref{cons}.

The constrained minimization allows us to construct an orbital free Hamiltonian, $H_\mathrm{OF}$, that produces the same self consistent potential and density as the TDDFT calculation. In this way the TDDFT
and the OF calculation can be directly compared.
\revcc{Since the constrained ground state density is the same, by setting  }
\begin{equation}
    V_c(\mathbf{r})=\lambda(\mathbf{r}),
\end{equation}
the contributions to the energy from the potentials are the same: 
\begin{equation}
    \langle \Psi\vert V_\mathrm{OF}\vert\Psi\rangle
    =2\sum_{i=1}^{N/2} 
    \langle\psi_i\vert V_{KS}\vert\psi_i\rangle.
\end{equation}
The kinetic energy parts in OF
\begin{equation}
    \langle \Psi\vert {\hbar^2\over 2m} \nabla_\mathbf{r}^2 \vert\Psi\rangle
\end{equation}
and in TDDFT 
\begin{equation}
    2\sum_{i=1}^{N/2} 
    \langle\psi_i\vert 
    {\hbar^2\over 2m} \nabla_\mathbf{r}^2
    \vert\psi_i\rangle
\end{equation}
are different, although in numerical calculations the difference is small. 
\revcc{The constraining potential forces to match the ground state OF density to the DFT density, and then applied the OF Hamiltonian. The constraining potential is kept fixed during the time evolution in these calculations.}

\subsection{\revc{The Riemann-Silberstein formalism}}
In a previous paper  \cite{PhysRevE.100.053301} we developed an approach to solve the coupled Maxwell TD-KS  equations in a numerically efficient way using the Riemann-Silberstein formalism.
\revc{The Riemann-Silberstein (RS) vector is defined as,
$$
\mathbf{F}(\mathbf{r}, t)=\sqrt{\frac{\epsilon_{0}}{2}} \mathbf{E}(\mathbf{r}, t) \pm i \sqrt{\frac{1}{2 \mu_{0}}} \mathbf{B}(\mathbf{r}, t),
$$
where $\mathbf{B}$ and $\mathbf{E}$ are magnetic and electric fields.} In this formalism the Maxwell equations can be rewritten in the form
$$
\nabla \cdot \mathbf{F}=\frac{1}{\sqrt{2 \epsilon_{0}}} \rho
$$
and
\begin{equation}
i \hbar \frac{\partial \mathbf{F}}{\partial t}=c\left(\mathbf{S} \cdot \frac{\hbar}{i} \nabla_\mathbf{r}\right) \mathbf{F}-\frac{i \hbar}{\sqrt{2 \epsilon_{0}}} \mathbf{J},
\label{rs}
\end{equation}
where $\mathbf{S}$ are $3\times 3$  the spin 1 Pauli matrices. The attractive feature of this formalism is that Eq. \eqref{rs} is similar to a time-dependent Schr\"odinger equation and can be solved with time propagation approaches used in quantum mechanics. 

To solve the Maxwell-equation \revcc{the electron density and current are calculated by solving
the TD-KS or TD-OF equations and then used to calculate $\mathbf{F}$ at time $t+\Delta t/2$.
Once $\mathbf{F}$ is known the vector potential, $\mathbf{A}$, at $t+\Delta t$ needed in the quantum equations can be calculated from $\mathbf{E}$ in a leapfrog algorithm. \cite{PhysRevE.100.053301}.}

\section{Results}
In this section we present test calculations comparing Maxwell-TDDFT and Maxwell-TD-OF calculations. Taylor time propagation is used to solve the TD-KS (Eq. \eqref{tdks}) and
TD-OF (Eq. \eqref{tdof}) equations. The time step in the time evolution of these equations
is $\Delta t=0.02$ a.u., and the time step in the propagation of the RS vector (Eq. 
\eqref{rs}) is $\Delta t/20$. The TD-KS and the TD-OF equations are solved on a numerical grid using a 9 point finite difference representation for the kinetic energy \cite{varga_driscoll_2011}. The grid spacing is $\Delta x=\Delta y=\Delta z=0.5$ for both jellium and for systems 
with atoms. The number of grid points is $N=N_x\times N_y \times N_z$, where $N_i$ is the number of grid points in the $i=x,y,z$ directions. The RS equations are solved in Fourier space \cite{PhysRevE.100.053301} corresponding to the same grid. The 
local density approximation is used for the exchange-correlation potential \cite{PhysRevB.23.5048} and the pseudopotential for Al is taken from Ref. 
\cite{B810407G}. Free or periodic boundary conditions (PBC)  are used in different directions as indicated in the examples. To avoid reflection of wave functions or electromagnetic waves complex absorbing potentials (CAP) are added at the boundary. The same CAP is used as in Ref. \cite{PhysRevE.100.053301}.   

\begin{figure}
\includegraphics[width=0.45\textwidth]{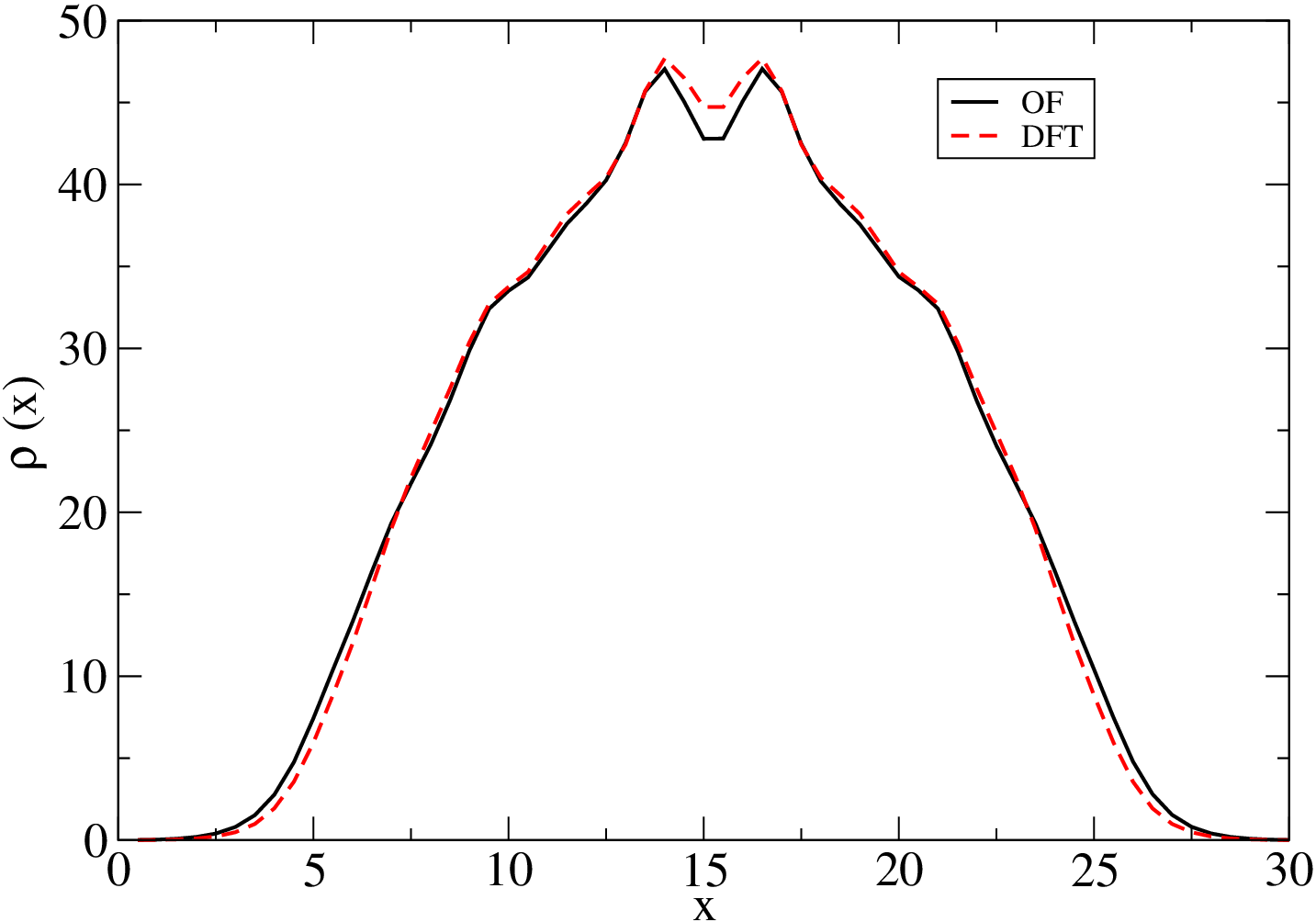}
\includegraphics[width=0.45\textwidth]{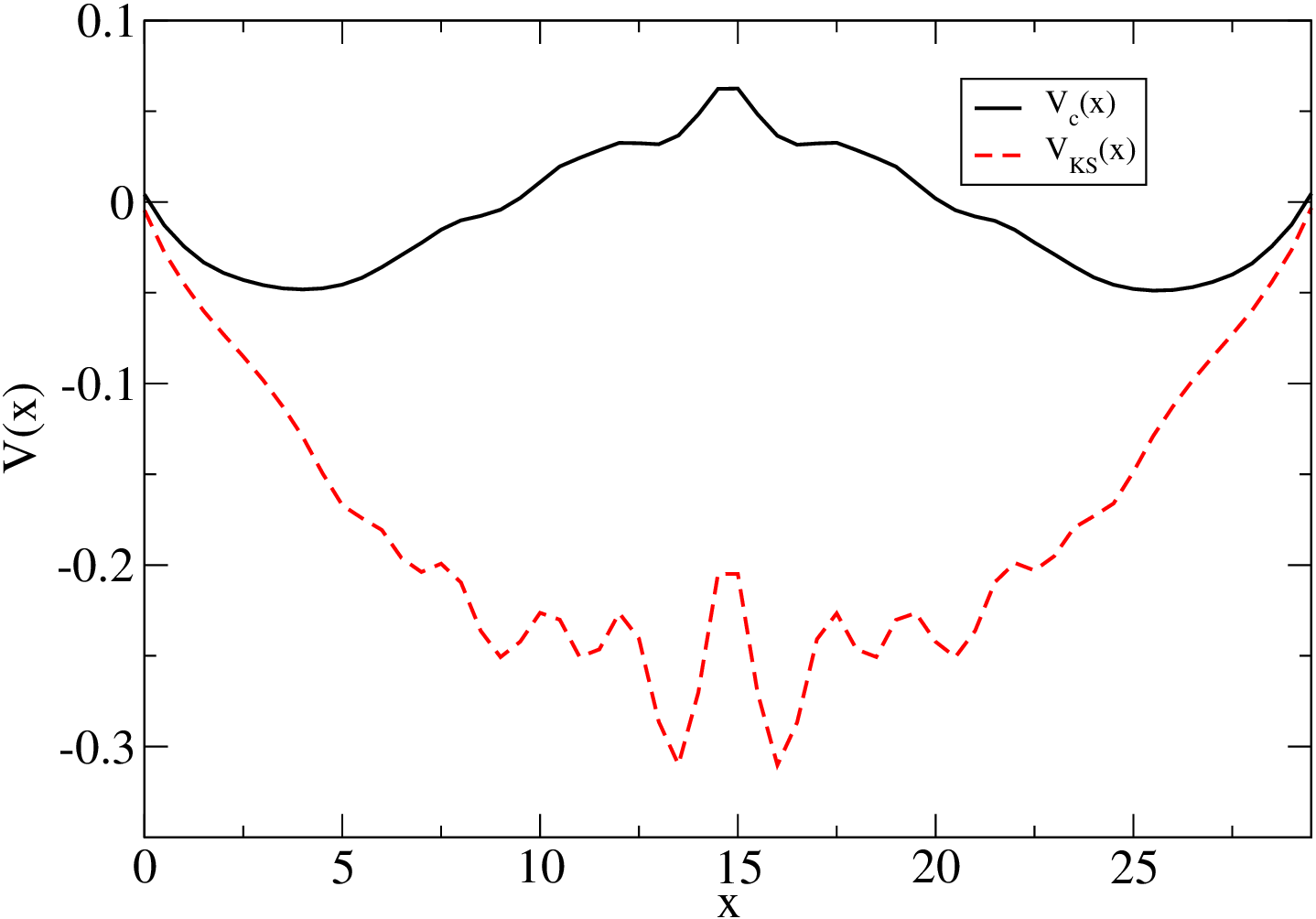}
\caption{Density and potential profile of an Al$_{55}$ cluster with 164 electrons. $N_x=N_y=N_z=60$ grid points used in the calculation. Top:
Comparison of the \revc{unconstrained--OF} and DFT electron densities averaged along the $x$ direction, $\rho(x)=\sum_{y,z}\rho(\mathbf{r})$. Bottom the confining and the KS potential averaged 
along the $x$ direction, $V(x)={1\over N_y N_z}\sum_{y,z} V(\mathbf{r})$.}
\label{cdft}
\end{figure}

The first step of the approach is a ground state DFT calculation to determine the initial wave function and initial density. Fig. \ref{cdft} compares the ground state density of the DFT and the OF calculation for an Al$_{55}$ icosahedron cluster with 165 electrons, using a geometry that is adapted from Ref. \cite{doi:10.1063/1.1574797}. 
\revc{To compare the ground state densities between DFT and the OF, the orbital free density was not constrained in this case, ie. $V_\mathrm{c}=0$.} The density of the DFT and the OF calculations are very close, but slight differences in the density cause a large difference in the potential (Fig.\ref{cdft} shows the magnitude of the necessary constraining potential) and in the energy. 

In the second step \revc{a the constrained minimization is used} to generate a potential, $V_\mathrm{c}$, that forces the OF calculation to produce the same density as $\rho$, the density obtained by the DFT
calculation. In this step we require 
\begin{equation}
    \vert\rho({\mathbf r})-\rho_\mathrm{OF}({\mathbf r})\vert < \epsilon,
\end{equation}
with $\epsilon=10^{-6}$. We start with $V_\mathrm{OF}$, $\Psi$, and $\rho_\mathrm{OF}$ calculated in the OF calculation setting $V_\mathrm{c}=0$. Then we set $\Psi^{(1)}=\Psi$,
$\lambda^{(1)}=0$ and use the iteration defined in eq. \ref{cdftiter} to calculate 
$\Psi^{(n+1)}$. The confining potential is updated in each step as $V_c(\mathbf{r})=\lambda^{(n)}(\mathbf{r})$.

\begin{figure}
\includegraphics[width=0.45\textwidth]{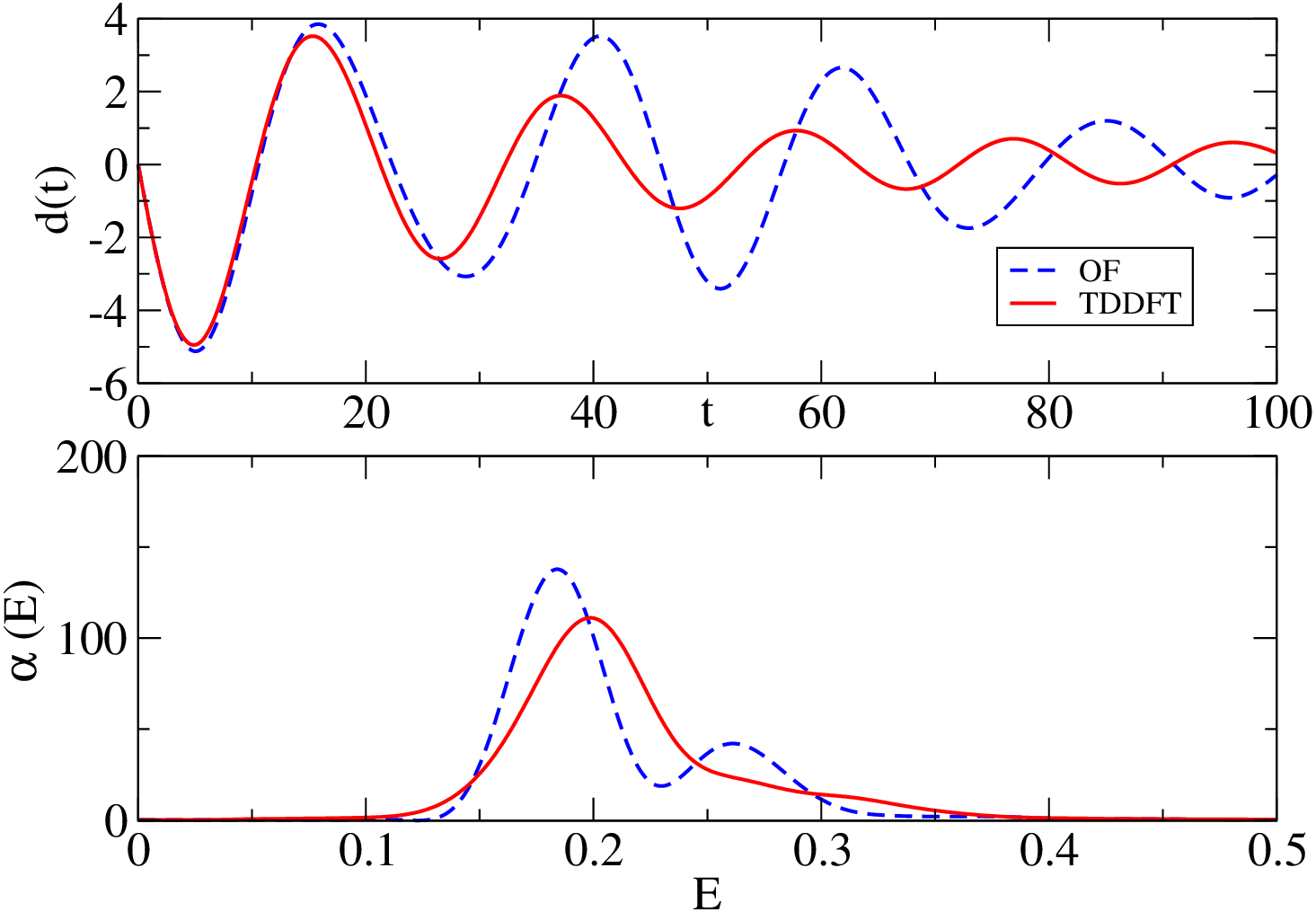}
\caption{Time dependence of the dipole moment and the absorption spectrum of an Al$_{55}$ cluster. Please see Fig. \ref{cdft} 
for details of the calculation. The absorption spectrum is defined as $\alpha(E)={e^2\over k\hbar}\int (d(t)-d(0))\mathrm{e}^{iEt/\hbar}g(t)dt$,
where $g(t)$ is a damping function.
} 
\label{abs}
\end{figure}

The calculated $V_\mathrm{c}$ is shown in Fig. \ref{cdft}. Using this constraining potential in $V_\mathrm{OF}$ one gets exactly the same self-consistent density in the OF and DFT calculations. The constraining potential is positive in the middle, decreasing the OF density to make it closer to DFT, and negative away from the center to pull the OF density closer to that of DFT.
To study the time-dependent behavior of the density in TDDFT and OF in this case we have calculated the 
time dependence of the dipole moment. To start the 
calculation the initial ground state wave functions were perturbed by multiplying them with
$\mathrm{e}^{ikx}$, with $k=0.01$ a.u. and the systems were time propagated up to $T=500$ a.u. The
calculated dipole moment $d(t)=\int \rho(\mathbf{r},t) x d\mathbf{r}$ is shown in Fig. 
\ref{abs}. The TDDFT and OF dipole moments are very 
close to each other in the beginning of the calculation, but later the oscillations are different. It seems that the constraining
potential works well in the initial stage of the time propagation, but later dynamical effects become 
important. This can probably be improved by invoking a dynamical kinetic energy potential
\cite{doi:10.1063/1.3574347,doi:10.1063/1.4867002}. The frequency of the oscillation of the dipole moment remains very similar in the 
OF and TDDFT calculation and the resulting absorption spectra are similar (see Fig. \ref{abs}). The usefulness of the OF approach 
for the calculation 
of the absorption spectrum of quantum dots using a dynamical kinetic energy functional \revcc{has been discussed} in Refs.  
\cite{doi:10.1021/jz500216t,doi:10.1021/acs.jpcc.6b05841}. 

\begin{figure}
\includegraphics[width=0.45\textwidth]{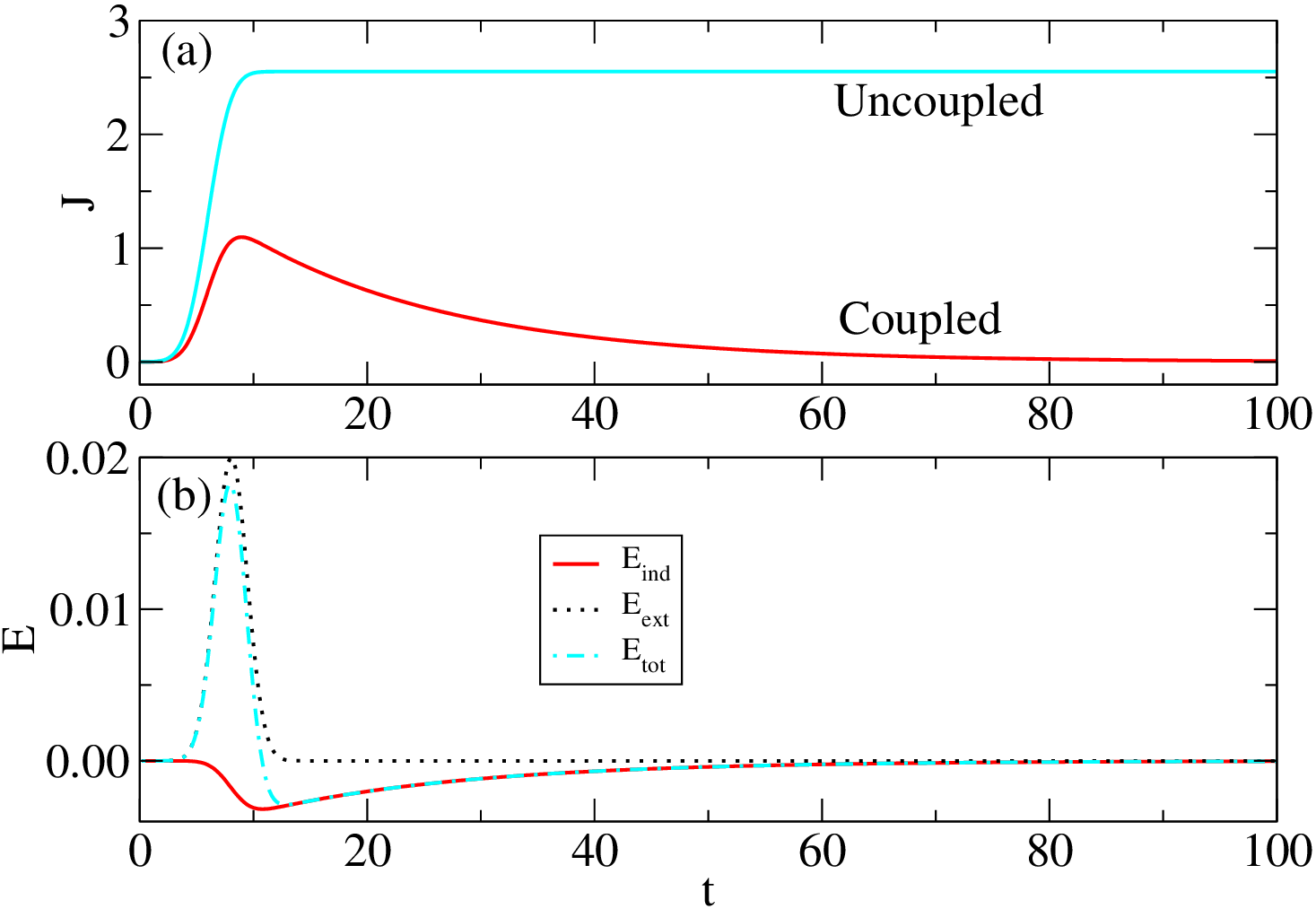}
\caption{Comparison of a TD-OF and TDDFT simulations showing the averaged currents 
(top) and the electric fields (bottom). The system is lithium jellium consisting of 36 electrons in a 328 a.u. wide sheet contained
within a box of $N_x$ = 1152, $N_y = N_z$ = 8. CAPs were used in the x direction and
PBCs were used in the y and z directions. The system was excited by
a Gaussian pulse with a peak electric field of $E_0$ = 0.02 a.u., a width
of $\alpha$ = 2.0 a.u. and a shift of $t_0$ = 6 a.u. The results of the TDDFT and TD-OF calculations agree up to 4 significant digits for the induced currents and induced fields and making the TDDFT and TD-OF calculations indistinguishable.}
\label{fig8}
\end{figure}

To test the coupled Maxwell TD-OF calculations we will compare them to coupled Maxwell-TDDFT calculations. 
First, we will use three test cases from Ref. \cite{PhysRevE.100.053301}. 
In the first case, an electromagnetic pulse excites a jellium sheet.
\revc{Because the Maxwell equations are linear, the electric field can be written as the sum of an external field and an induced field created by the electron currents, $\mathbf{E}_{tot} =\mathbf{E}_{ext} + \mathbf{E}_{ind}$. In this way, the exciting pulse does not need a source and never has to be time propagated because its form is analytic}. 
The \revc{external field is given} as a Gaussian laser pulse ($\mathbf{E}_{ext}$ in Fig. \ref{fig8})
\begin{equation}
\mathbf{E}_{ext}(\mathbf{r}, t)=(0,0,E_{0} e^{-\left(t-t_{0}-x / c\right)^{2} /\alpha^2}).
\end{equation}
Here $t_{0}$ is the pulse shift, $\alpha$ controls the width of the pulse. 

To show the time-dependence of the electric field and currents we define the average electric field
as
\begin{equation}
    \mathbf{E}(t)={1\over N} \sum_\mathbf{r} \mathbf{E}(\mathbf{r},t),
\end{equation}
and 
\begin{equation}
    \mathbf{J}(t)={1\over N} \sum_\mathbf{r} \mathbf{J}(\mathbf{r},t),
\end{equation}
where the  sum is over grid points that are not in the CAP region. In  the figures only the $z$
components $E=E_z$ (for $\mathbf{E}_{tot}, \mathbf{E}_{ext}$ and $\mathbf{E}_{ind}$), and $J=J_z$ are shown.

Fig.\ref{fig8}a shows the induced currents  for coupled and uncoupled cases.
\revc{In the uncoupled case, the Maxwell equations are not time propagated and $\mathbf{E}_{tot} = \mathbf{E}_{ext}$. For the coupled case, the Maxwell equations are time propagated and the currents produce an induced field opposite to $\mathbf{E}_{ext}$. It is this induced field that de-excites the electrons and causes the current to fall back to zero.}
The results of the TDDFT and TD-OF calculations agree up to 4 significant digits for the induced currents and induced fields and one cannot distinguish the results on Fig. \ref{fig8}.  

The perfect agreement is very surprising considering that 18 orbitals are time propagated in TDDFT and only one in TD-OF. We will see later that this agreement is mostly due to symmetry of the jellium sheet system. Fig.\ref{fig8}.a also shows that in the uncoupled case the system stays in an excited state while in the coupled case the system relaxes to the ground state and the electric current decreases to zero. 

In the second test case a sheet of aluminium jellium is irradiated with a laser pulse \revcc{of the form}
\begin{equation}
\mathbf{E}_{ext}(\mathbf{r}, t)=(0,0,E_{0} e^{-\left(t-t_{0}-x / c\right)^{2} /\alpha^2}\mathrm{sin}(\omega(t-x/c))).
\end{equation}
The TDDFT and OF currents and induced electric fields are in perfect agreement, up to 3 significant digits. In this case the electric field penetrates the thick Al sheet generating 
strong currents. Still, the results of the TDDFT calculations with 200 orbitals and the orbital
free calculation are the same. 
\begin{figure}
\includegraphics[width=0.45\textwidth]{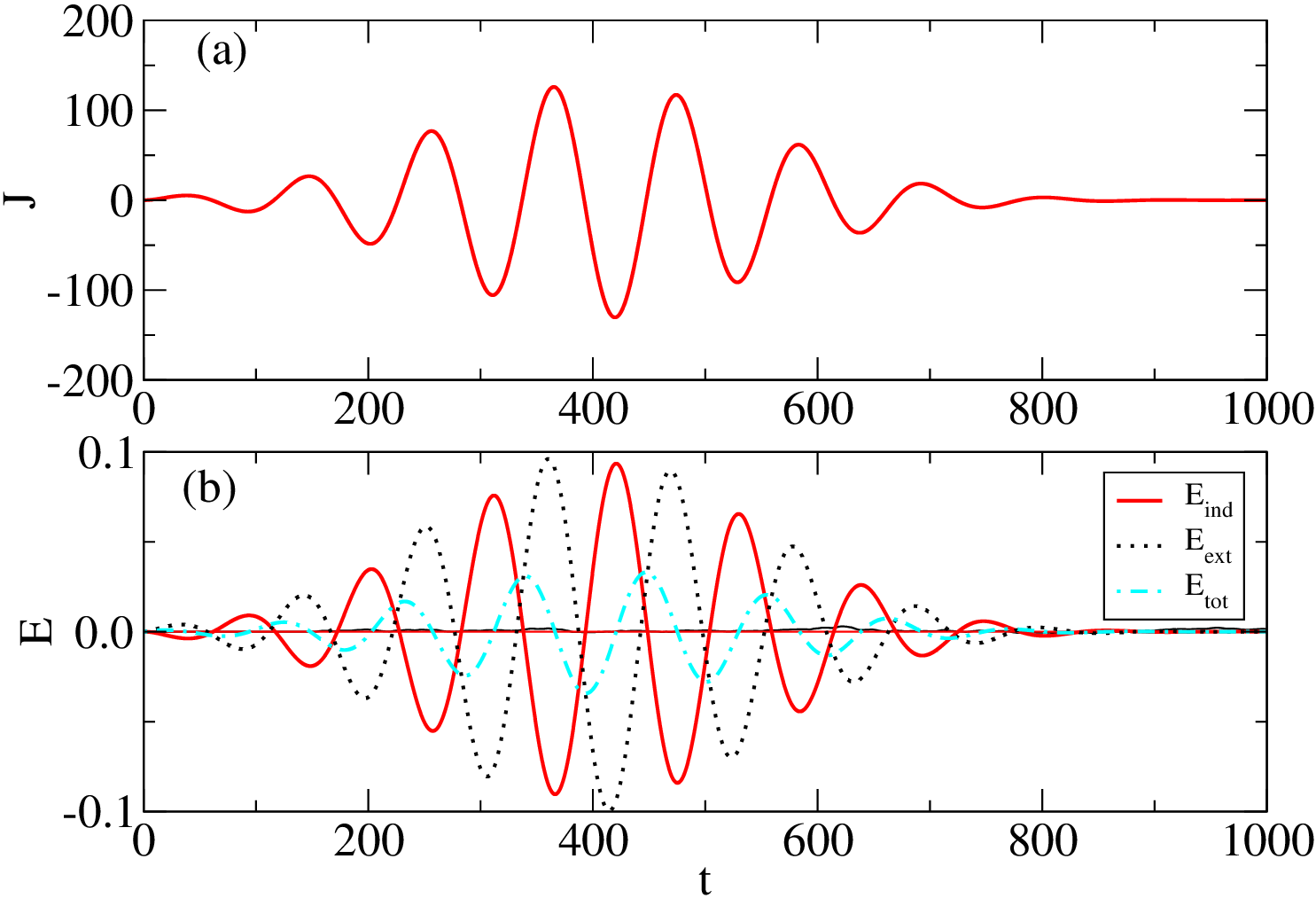}
\caption{Calculated currents and induced fields in the case of a laser pulse
incident from the left on a 232 a.u. thick sheet of aluminum-jellium
of 400 electrons within a box of $N_x = 640, N_y = N_z = 16$. \revcc{Also in this case the OF and TDDFT curves are on top of each other}. 
CAPs were used in the x direction and PBCs were used in the y and z directions. The
system was excited by a Gaussian pulse with a peak electric field
of $E_0$ = 0.1 a.u., a width of $\alpha$ = 206 a.u., $\omega$=0.05695 a.u., and a
shift of $t_0$ = 400 a.u..}
\label{fig9}
\end{figure}

\begin{figure}
\includegraphics[width=0.45\textwidth]{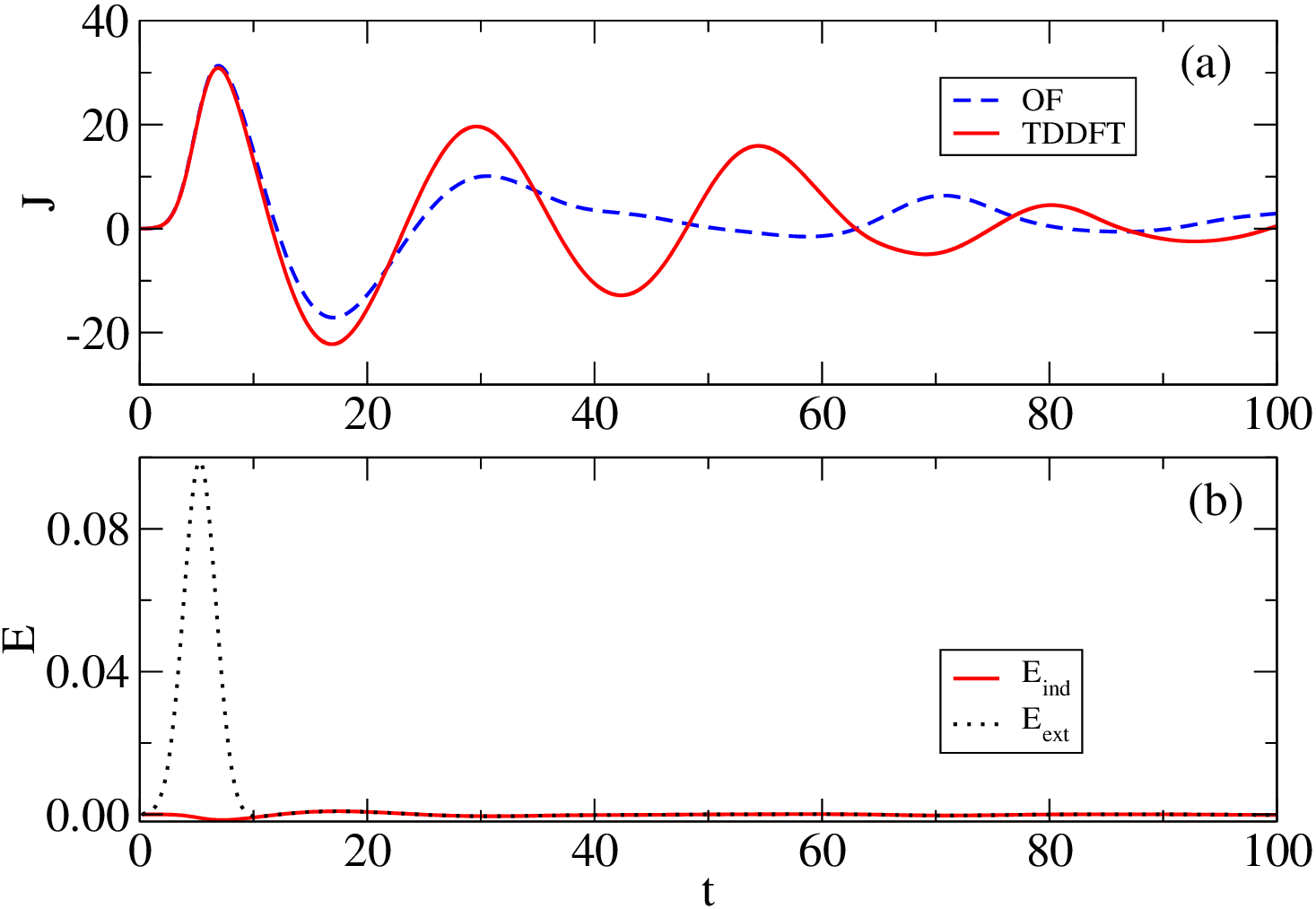}
\caption{Comparison of the cell average currents in time resulting 
from a strong laser pulse incident from the left on a sphere
of aluminum-jellium of 236 electrons in a box of $N_x = 192 a.u.,
N_y = N_z = 60.$ a.u.
CAPs were used in the x direction and PBCs were used in the y and z
directions. The system was excited by a Gaussian pulse with a peak
electric field of $E_0$ = 0.1 a.u. with a width of $\alpha$ = 2 a.u. and a shift
of $t_0$ = 5 a.u..}
\label{fig11}
\end{figure}

\begin{figure}
\includegraphics[width=0.45\textwidth]{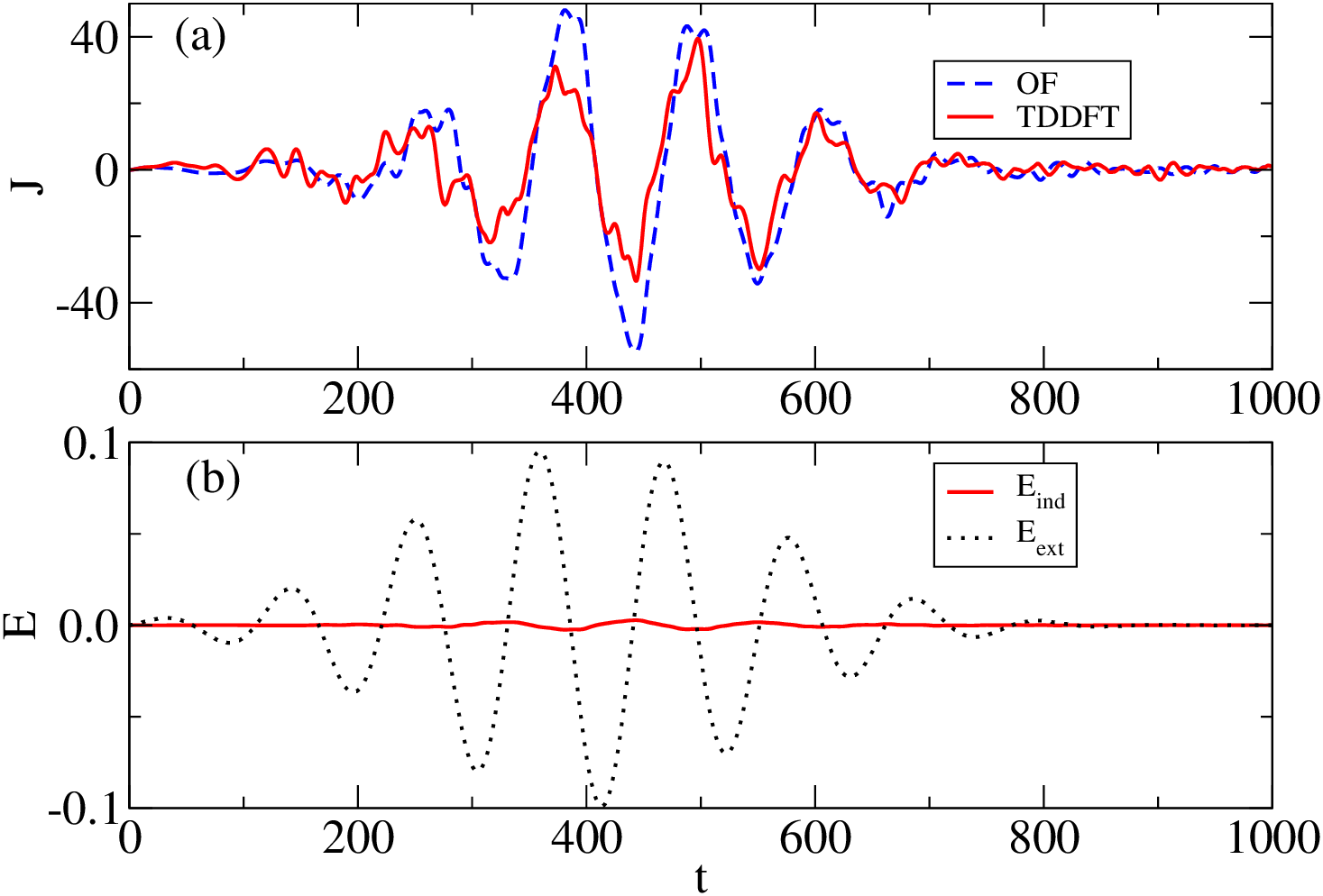}
\caption{The same system as described in the caption of Fig. \ref{fig11} but with the laser pulse of Fig. \ref{fig9}.}
\label{fig11_long}
\end{figure}

In the third example, a jellium sphere of radius 12.8 a.u. containing 236 electrons is used with a short Gaussian pulse (see Fig. \ref{fig11}). In this case the induced electric field 
is negligible. The OF and TDDFT currents are very close to each other during the laser pulse, but after the excitation the time dependence of the currents is different. The same system
with the longer pulse is shown in Fig. \ref{fig11_long}. The OF and TDDFT are not in perfect agreement as before, but they are similar and they follow the external pulse with the same delay time. This test shows the role of symmetry comparing a sheet and a spherical jellium system. One can cut the sphere into $n$ slices in parallel with the direction of the laser.
The currents and density oscillations will be different in each slice depending on their diameter.  In the jellium sheet cases the response is the same in each slice.
\revcc{Additionally, the jellium sphere presents a surface which causes reflections of the electron density driven to the surface by the laser field. 
In the case of TDDFT, there are 118 orbitals with different energies and different reflections. For larger
systems the effect of the surface is expected to be smaller.}

\begin{figure}
\includegraphics[width=0.45\textwidth]{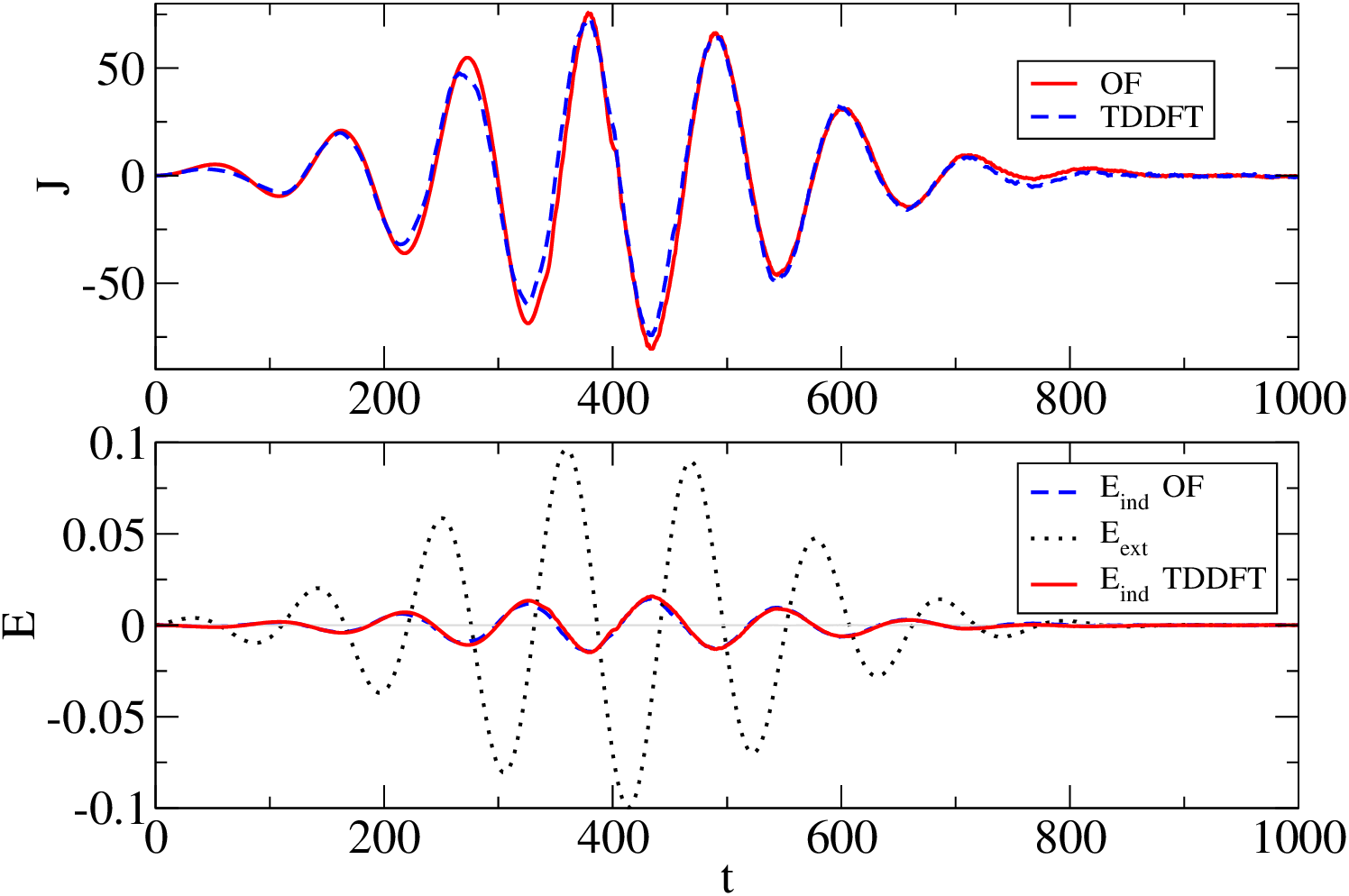}
\caption{Comparison of OF and TDDFT currents and induced electric fields in a 36 atom,  96 electron Al sheet. $N_x=200,N_y=N_z=31, \Delta_x=\Delta_y=\Delta_z$=0.4932 a.u. is used. 
A CAP is added in the x direction and PBCs in the y and z directions. The parameters of the laser field are given in the caption of Fig. \ref{fig9}.}
\label{Al32_long}
\end{figure}
\begin{figure}
\includegraphics[width=0.45\textwidth]{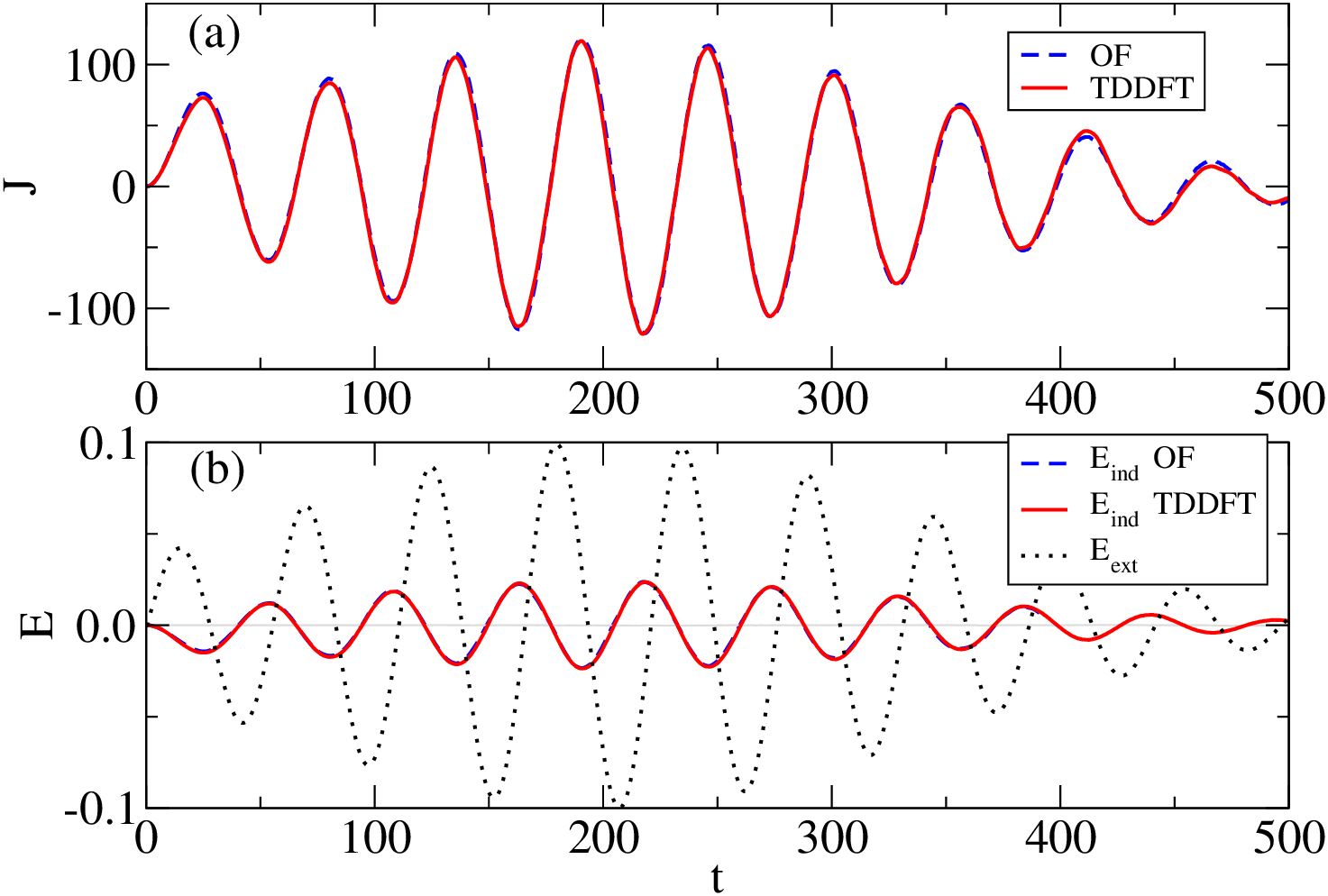}
\caption{Comparison of OF and TDDFT currents and induced electric fields in a 96 atom  288 electron Al sheet. $N_x=200, N_y=N_z=31, \Delta_x=\Delta_y=\Delta_z$=0.4932 a.u. is used. 
A CAP is added in the x direction and PBCs in the y and z directions.
The laser pulse have a peak electric field
of $E_0$ = 0.1 a.u., a width of $\alpha$ = 200 a.u., $\omega$=0.1139 a.u., and a
shift of $t_0$ = 200 a.u..}
\label{Al96_long}
\end{figure}

Next we present examples with  atoms. In Figs. \ref{Al32_long} and \ref{Al96_long} OF and TDDFT
calculations of a 32 atom and 96 atom Al sheets are compared. The agreement of the OF and TDDFT results is excellent, the thicker (the width is about 42 a.u.) sheet of 96 atom system has better agreement than the thinner case. This example illustrates that the presence of atomic potentials does not greatly affect the agreement between the OF and TDDFT results.

Finally, we present a calculation for a Al$_{147}^+$ icosahedron cluster. As the TDDFT calculations are computationally demanding, we use a shorter pulse in this case. The coupling 
to the Maxwell equations is also turned off -- partly to increase the speed of the calculation partly because the induced fields are small as we have seen in the jellium case (see Fig. \ref{fig11_long}. The agreement of the TDDFT and OF calculations is perfect. The reason of the very good agreement is probable due to the localization of the density by the atomic potentials. In the jellium case, the electron density \revcc{is more easily moved by the field.}

\section{Summary}
We have implemented and tested the coupled Maxwell and TD-OF calculations to study the interaction
of electromagnetic fields and matter. Nanometer \revcc{sized} sheets and clusters were subject to short laser pulses and the induced currents and electric fields were compared to the result of TDDFT
calculations. The results are in very good agreement, especially for larger systems. 
The examples and tests include dipole oscillation due to an instantaneous kick perturbation,
excitation with a laser field and coupled Maxwell and quantum dynamics.
\revcc{The coupled Maxwell case is a rigorous test because the system does not just follow the dynamics of the laser field
but builds up a nonlinear response. This test is also important for possible applications.}

The  OF calculations have some limitations for smaller quantum dots \cite{HO200757,PhysRevB.63.115404,doi:10.1021/jp0018504} and the disagreement can perhaps be 
reduced by using more recently proposed kinetic energy functionals \cite{PhysRevB.100.041105,shao2020newgeneration}. 
\revcc{In these calculations, a constraining potential was used to force the OF ground state density to match the DFT ground state density.}
This step can probably be eliminated by using new generations of kinetic energy functionals 
\cite{PhysRevB.100.041105,shao2020newgeneration,doi:10.1063/1.5023926,PhysRevB.100.205132,PhysRevB.100.041105,PhysRevB.101.045110}. 

Another potential improvement is using  dynamic kinetic energy potentials \cite{doi:10.1063/1.3574347,PhysRevB.98.144302}. These potentials have been tested only in very few cases and further development and tests might be needed.

In the present work the size of the studied systems is relatively small because TDDFT calculations are not feasible for larger systems. The results are promising and larger, physically more relevant systems will be studied in later works.


\begin{figure}
\includegraphics[width=0.45\textwidth]{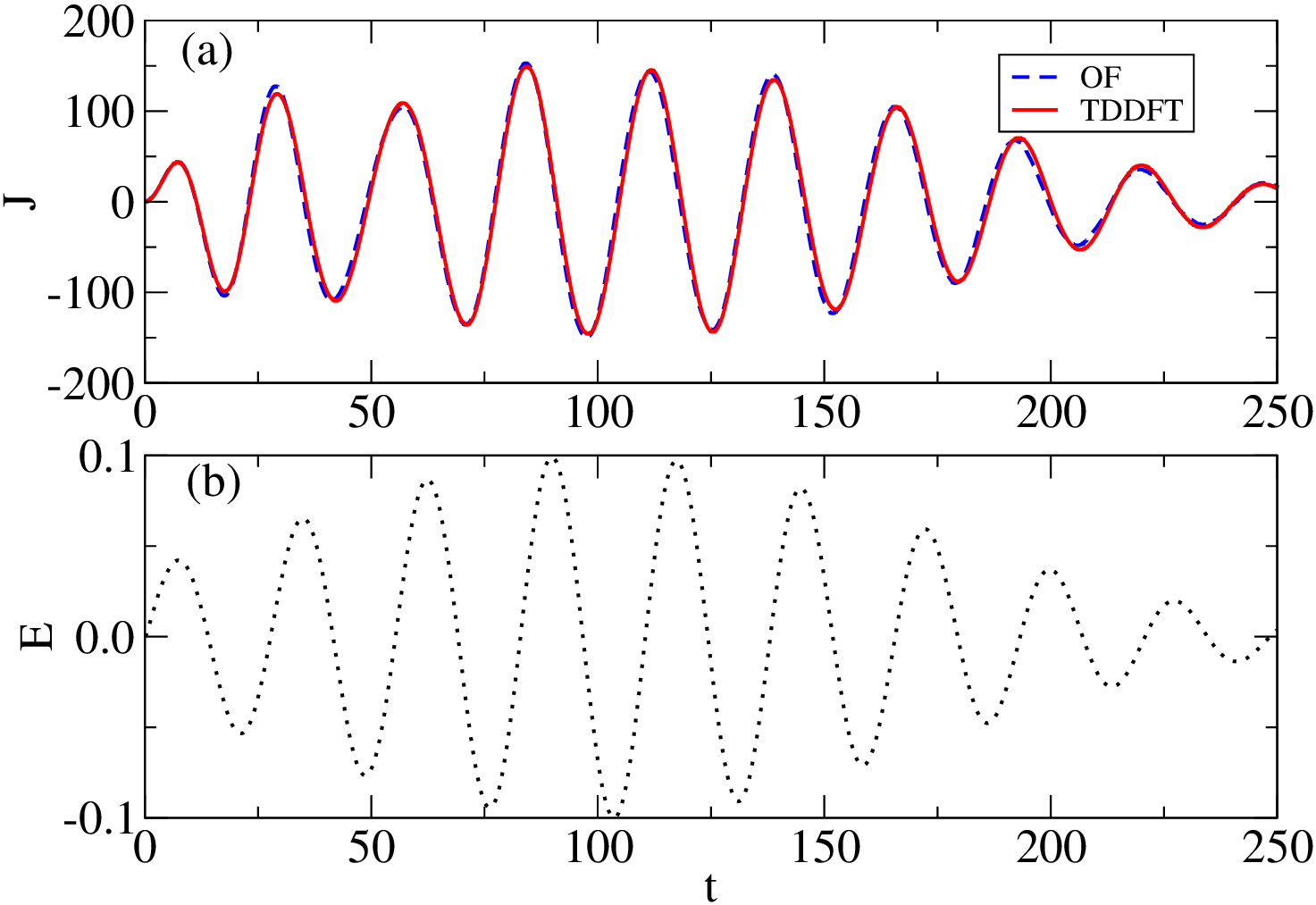}
\caption{Comparison of OF and TDDFT currents  in a 147 atom icosahedron Al cluster with  440 electrons. The number of grid points is $N_x=200,N_y=N_z=60$, and a  CAP is added in the x direction and free boundary conditions are used  in the y and z directions. The laser pulse have a peak electric field
of $E_0$ = 0.1 a.u., a width of $\alpha$ = 100 a.u., $\omega$=0.2278 a.u., and a
shift of $t_0$ = 100 a.u..}
\label{Al147_long}
\end{figure}

%


\end{document}